\documentclass{article}
\usepackage{mathtools}
\usepackage{enumitem}
\usepackage{spconf,amsmath,graphicx}
\usepackage{subcaption} 
\usepackage{amssymb}
\usepackage{epsfig}
\usepackage{caption3}
\usepackage{acronym}
\usepackage{fancyhdr}
\usepackage{color}
\usepackage{verbatim}
\usepackage{listings}
\usepackage{graphicx}
\usepackage{cite}
\usepackage{caption}
\usepackage{hyperref}
\usepackage{subcaption}
\usepackage{multirow}

\usepackage{soul, color}
\sethlcolor{yellow}

\def\n{{\mathbf n}}

\def\0{{\mathbf 0}}
\def\1{{\mathbf 1}}

\title{IMPROVING TRAJECTORY LOCALIZATION ACCURACY VIA DIRECTION-OF-ARRIVAL DERIVATIVE ESTIMATION}

\name{Ruchi Pandey$^\dagger$, Shreyas Jaiswal$^\dagger$, Huy Phan$^*$, and  
Santosh Nannuru$^\dagger$}
\address{$^\dagger$ Signal Processing and Communication Research Centre, IIIT Hyderabad, India, \\
	$^*$ School of Electronic Engineering and Computer Science, Queen Mary University of London, UK}
\begin{document}
%
\maketitle
\begin{abstract}


Sound source localization is crucial in acoustic sensing and monitoring-related applications. In this paper, we do a comprehensive analysis of improvement in sound source localization by combining the direction of arrivals (DOAs) with their derivatives which quantify the changes in the positions of sources over time. This study uses the SALSA-Lite feature with a convolutional recurrent neural network (CRNN) model for predicting DOAs and their first-order derivatives. An update rule is introduced to combine the predicted DOAs with the estimated derivatives to obtain the final DOAs. The experimental validation is done using TAU-NIGENS Spatial Sound Events (TNSSE) 2021 dataset. We compare the performance of the networks predicting DOAs with derivative vs. the one predicting only the DOAs at low SNR levels.
The results show that combining the derivatives with the DOAs improves the localization accuracy of moving sources.

\end{abstract}

\vspace{-0.2cm}
\section{Introduction}
Sound source localization is an integral part of many modern applications such as video conferencing, hearing aids, and human-robot interactions \cite{farmani2015maximum,wan2016application}. There are a plethora of localization methods existing in the literature \cite{vantrees2002,music1986,gccphatanalysis,compressivebeamforming,gerstoft2016,pandey2021sparse}. Recently data-driven methods have shown promising results for sound source localization in reverberant and low signal-to-noise ratio (SNR) scenarios \cite{diaz2020robust,opochinsky2021deep,adavanne2018sound,grumiaux2022survey}.        
The deep neural network (DNN) based methods were shown to outperform parametric methods in terms of high resolution and low erroneous DOAs \cite{chakrabarty2017broadband,adavanne2019multi,xiao2015learning,suvorov2018deep,grumiaux2022survey,nguyen2018autonomous}. 

In the recent past, polyphonic sound event localization and detection (SELD) problems have garnered a lot of attention among researchers, which combine the detection and localization tasks and have many practical applications \cite{adavanne2018direction,adavanne2018sound,adavanne2019multi}. Since its introduction, various model architectures and features have been proposed \cite{shimada2021accdoa,phan2020multitask,cao2021improved,nguyen2022salsa,2022salsa-lite}. In this study, we focus on improving the localization accuracy of an existing model \cite{2022salsa-lite} by predicting both the DOAs and their derivatives, i.e., changes in DOAs over time. In the real world, the sources move non-linearly and non-uniformly; hence estimating the derivatives potentially improves the model learning resulting in accurate source trajectories. We compare the existing localization models (considering the detection ground truth to be known), which predicts only DOAs, with the model, which predicts both DOAs and their derivatives.

Our experiments reveal that localization can be a challenging task, even for immobile sound sources.
This research aims to better understand deep learning-based models for localization task with a special focus on combining DOAs with their derivatives to improve the source trajectories. We hope the analysis will direct the research focus towards localization aspect of SELD task.

The specific contributions of this paper are \\
{\bf{(a)}} deriving an update rule incorporating predicted DOAs and derivatives to improve localization accuracy; \\
{\bf{(b)}} conduct experimental validation using TAU-NIGENS Spatial Sound Events 2021 dataset \cite{politis2021dataset}; \\
{\bf{(c)}} carry out performance analysis of the proposed model at different SNR levels by synthetically adding noise; \\
{\bf{(d)}} performance analysis of the proposed model when using the pre-trained model for initialization.
\vspace{-0.3cm}

\section{Model architecture }
\label{sec:architecture}
\subsection{Features}
\label{sec:salsa-lite}
The SALSA-Lite was introduced as an efficient computational version of the Spatial Cue-Augmented Log-Spectrogram (SALSA) feature for MIC (audio format) data \cite{nguyen2022salsa,2022salsa-lite}. For $M$-channel audio recording, SALSA-Lite is a ($2M-1$) channel feature consisting $M$ log-power spectrogram with ($M-1$) frequency-normalized interchannel phase differences (NIPDs). 
The NIPD ($\Lambda$) approximating the relative distance of arrival (RDOA) can be written as      
\begin{align}
    \Lambda(t,f) &\approx -\frac{c}{2\pi f}\, \text{arg} ~|\boldsymbol{H}_1^*(t,f)\boldsymbol{H}_{2:M}(t,f)|, \\
    &\approx [d_{12}(t)\hdots d_{1M}(t)],
\end{align}
where $\boldsymbol{H}_m(t,f) = e^\frac{j2\pi f d_{1m}(t)}{c}$ is the array response for any arbitrary array structure under the far-field assumption and $d_{1m}(t)$ is the RDOA between the first (reference) and $m^{\text{th}}$ mic.  The SALSA-Lite provides the exact time-frequency positioning between the spectrogram and the NIPD resulting in the model being able to localize multiple overlapping sources.
\vspace{-0.3cm}
\subsection{Architecture}
\begin{figure}[h]
    \centering
	\includegraphics[width=0.44\textwidth]{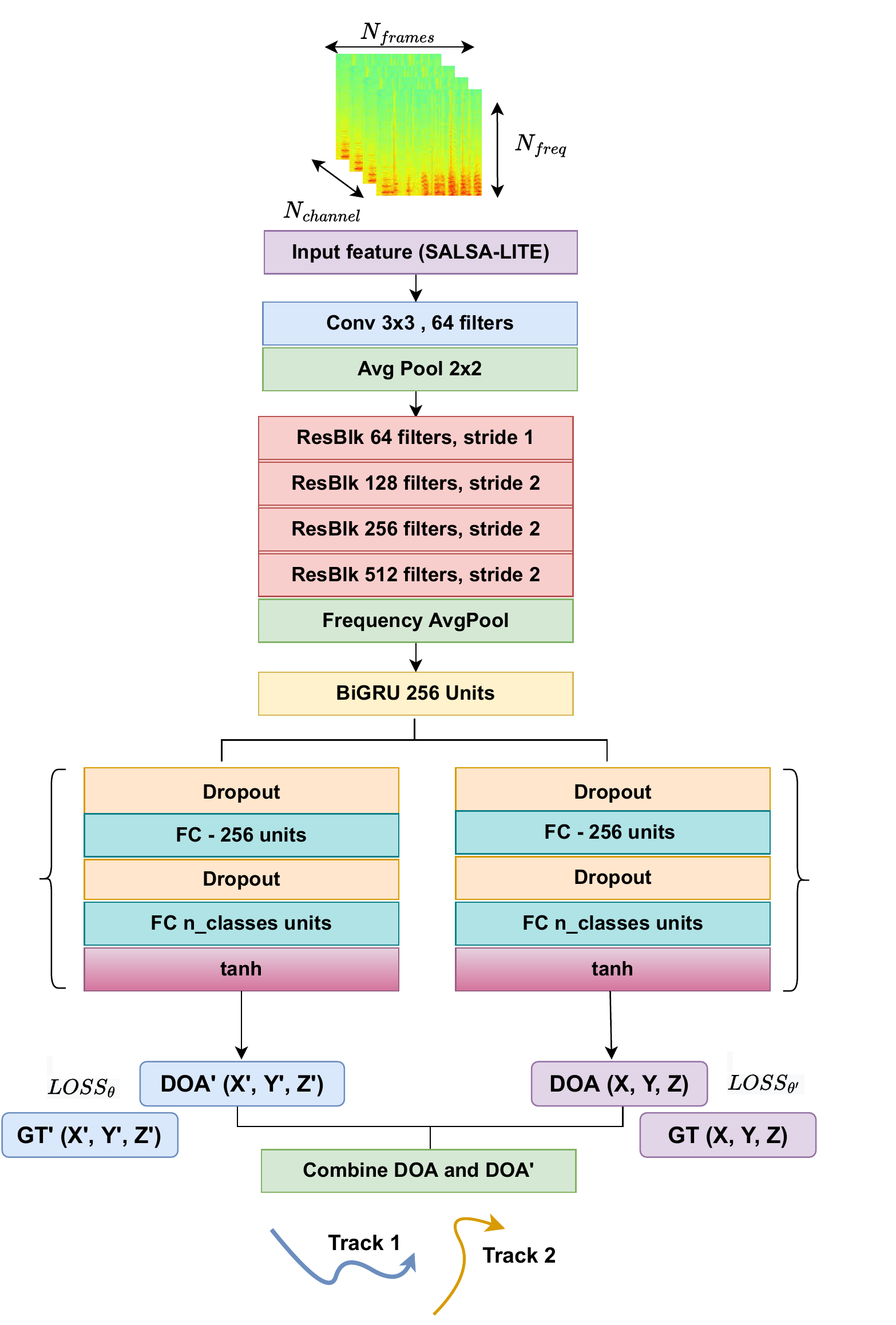}
	\caption{Model architecture predicting both DOAs and DOA derivatives}
	\label{nn}
	\vspace{-0.2cm}
\end{figure}
Figure \ref{nn} shows the neural network architecture designed to predict both the DOAs and their derivatives simultaneously. Here derivative represents the change in DOAs between two time frames. The SALSA-lite is fed to the CRNN network, which consists of one convolutional layer followed by four ResNet22 blocks \cite{kong2020panns} in the network body \cite{nguyen2022salsa,2022salsa-lite}. 


The output of ResNet block is fed into a two-layer bidirectional Gated Recurrent Unit (GRU) followed by two distinct regression heads for predicting DOAs and their derivatives in Cartesian coordinates (X, Y, and Z), respectively. Unlike the SELD network in \cite{2022salsa-lite}, we focus only on the localization task and replace the detection head with the regression head, which predicts the derivatives of DOAs at different timestamps as shown in Fig. \ref{nn}. Along with predicting the intermediate DOAs, the additional derivative information helps to obtain better overall DOA estimates. Once the network predicts the DOAs and their derivatives, the final DOAs are obtained using the following update equation
\begin{align}
   \hat{y}_N^{\text{final}}~ = ~\frac{\hat{y}_N~+~(\hat{y}_{N-1}~ +~ \hat{y}'_{N})}{2},
   \label{eq:update}
\end{align}
where $\hat{y}'_N$ is an estimate of the DOA derivative $(y_N - y_{N-1})$ at $N^{\text{th}}$ time and $\hat{y}'_0 = 0$ is the first derivative assumed to be zero.  
We expect that by additionally incorporating derivative predictions, DOA of moving targets can be estimated more accurately. For both the regression heads (predicting DOAs and derivatives), the number of active sources are assumed to be known. The ground truth is used to compute the losses for both the prediction heads. The mean squared error (MSE) loss is minimized while training both the network heads and can be written as
\begin{equation}
\text{LOSS}_{\text{total}} =    \sum_n (y - \hat{y})^2  + \sum_n (y' - \hat{y}')^2,
\end{equation}

where $y ~\text{and}~y'$ are true DOAs and their derivatives at different timestamps, and $\hat{y}$ and $\hat{y}'$ represent the predicted DOAs and the estimated derivatives, respectively.
\vspace{-0.2cm}
\section{Simulations and results}
\label{sec:results}
\subsection{Dataset and training}
The TAU-NIGENS Spatial Sound Events 2021 dataset was used for the performance analysis of the proposed models. The dataset consists of 600 recordings (1 minute long and four channels each) with 24 kHz sampling frequency in MIC data format. In this paper, 400 recordings were used for training, and 100 recordings were used each for validation and testing. The audio files consist of various static and moving sources from 12 unique classes. The angular range for azimuth and elevation angles are $[-180 ~180)^\circ$ and $[-45~ 45]^\circ$, respectively. 

For feature extraction, we followed the same setup as in \cite{2022salsa-lite}, and processed frequencies from 50Hz to 2kHz to avoid the aliasing. While training, Adam optimizer \cite{kingma_2014} was used, with the initial learning rate as $3 \text{x} 10^{-4}$ which linearly decreased to $10^{-4}$ over last 15 epochs. A total of 70 epochs with $32$ batch size were processed. The validation set was used for model selection, whereas the test data was used for the performance analysis. While generating the ground truth for DOA derivatives, the derivative is considered zero if the source appears for the first time or reappears after 20 or more frames.
In all other cases the derivative is the difference between the previous and current DOA.


\vspace{-0.2cm}
\subsection{Accuracy metric}
Following the framework of baseline method \cite{2022salsa-lite}, the detection ground truth was used to compute the error only for the frames where sources were present. For this study, the error was computed as the DOA error $\Delta\sigma$ which corresponds to the spatial angular distance between the predicted and true positions, \cite{mesaros2019joint,adavanne2019multi}. 
\begin{equation}
\Delta\sigma = \arccos{(\n_{\text{true}}. \n_{\text{pred}})}\,.\, \frac{180}{\pi}, 
\end{equation}
where $\n_{\text{true}}$ and $\n_{\text{pred}}$ are unit norm vectors of [X, Y, Z] coordinates corresponding to the true and predicted positions, respectively. A source is considered to be localized only when the average (averaged across the active frames) spatial distance between the predicted and true positions is less than $20^\circ$; we call these cases as true positive (TP). The false negative (FN) counts the number of incidences when the averaged spatial distance is more than $20^\circ$. The probability of detection ($P_d$) is the fraction of frames where the distance between the predicted and true positions is less than $20^\circ$. Note that as the network is designed to provide only one prediction per class, the error was not calculated for the cases where multiple sources were present from the same class in the frame.

\vspace{-0.3cm}
\subsection{Effect of combining derivative}
\begin{figure*}
    \centering
	\includegraphics[width=0.8\textwidth]{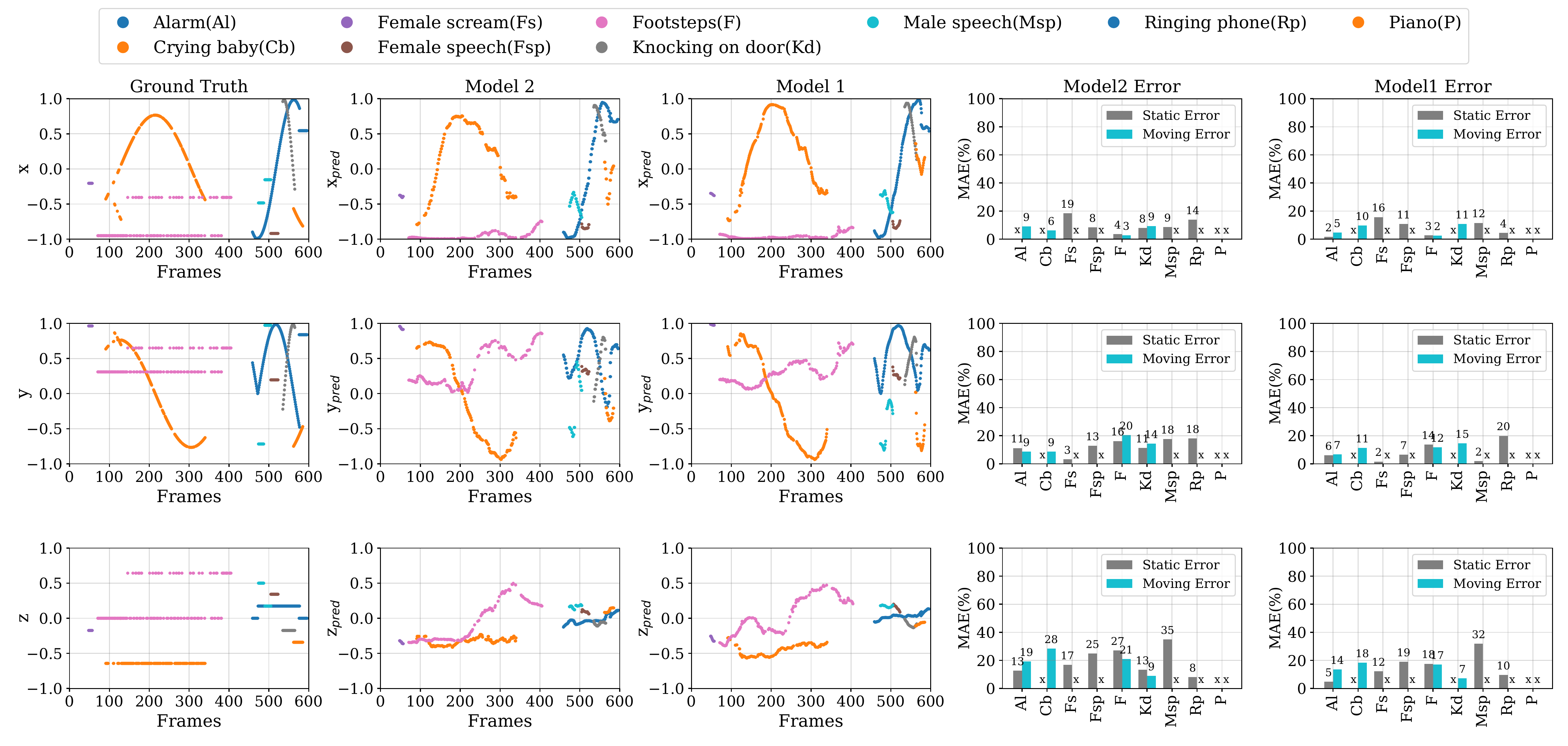}
		\vspace{-0.2cm}
	\caption{Effect of derivatives: true and predicted trajectories from Model1 and Model2 with classwise MAE.}
	\label{fig:grad_effect}
\end{figure*}

In this subsection, the effect of combining the predicted derivatives with predicted DOAs is demonstrated. Fig. \ref{fig:grad_effect} shows the predicted source trajectories from both Model1 (with derivative estimation) and Model2 (without derivative estimation) along with classwise mean absolute error (MAE) computed for the correctly detected sources for one of the recordings. The cross $(\times)$ in MAE plot denotes the cases when the source has not been detected by the models. It can be seen that Model1 detects more number of sources, hence resulting in higher MAE for some classes compared to Model2. It can be observed that by combining the derivatives via the proposed update rule \eqref{eq:update}, the final DOA exhibits a smoother trajectory since the outliers are eliminated. The update rule is helpful even for static sources. We observed that Model2 gives more erroneous DOAs for static sources than Model1. Overall, Model1's estimates are closer to true trajectories resulting in higher $P_d$. For this recording, the average $P_d$ for static and moving sources for Model1 and Model2 are, $P_{ds} = 64$, $P_{dm} = 78$, $P_{ds} = 53$, and $P_{dm} = 52.4$, respectively. The total $P_d$ averaged over 100 recordings from the test data is reported in Table \ref{tab:SNRs}. 

From Table \ref{tab:SNRs}, it is evident that both Model1 and Model2 show similar performance for the clean dataset. We observed that Model1's performance degrades when the network predicts the erroneous DOAs; hence combining them with equal weights leads to incorrect estimates. As a correction step, the current DOA prediction with derivative and the past prediction can be weighted depending on the threshold. A choice must be made depending on confidence in the present and past predictions.
\vspace{-0.3cm}

\subsection{Effect of transfer learning}
\begin{figure*}
    \centering
	\includegraphics[width=0.8\textwidth]{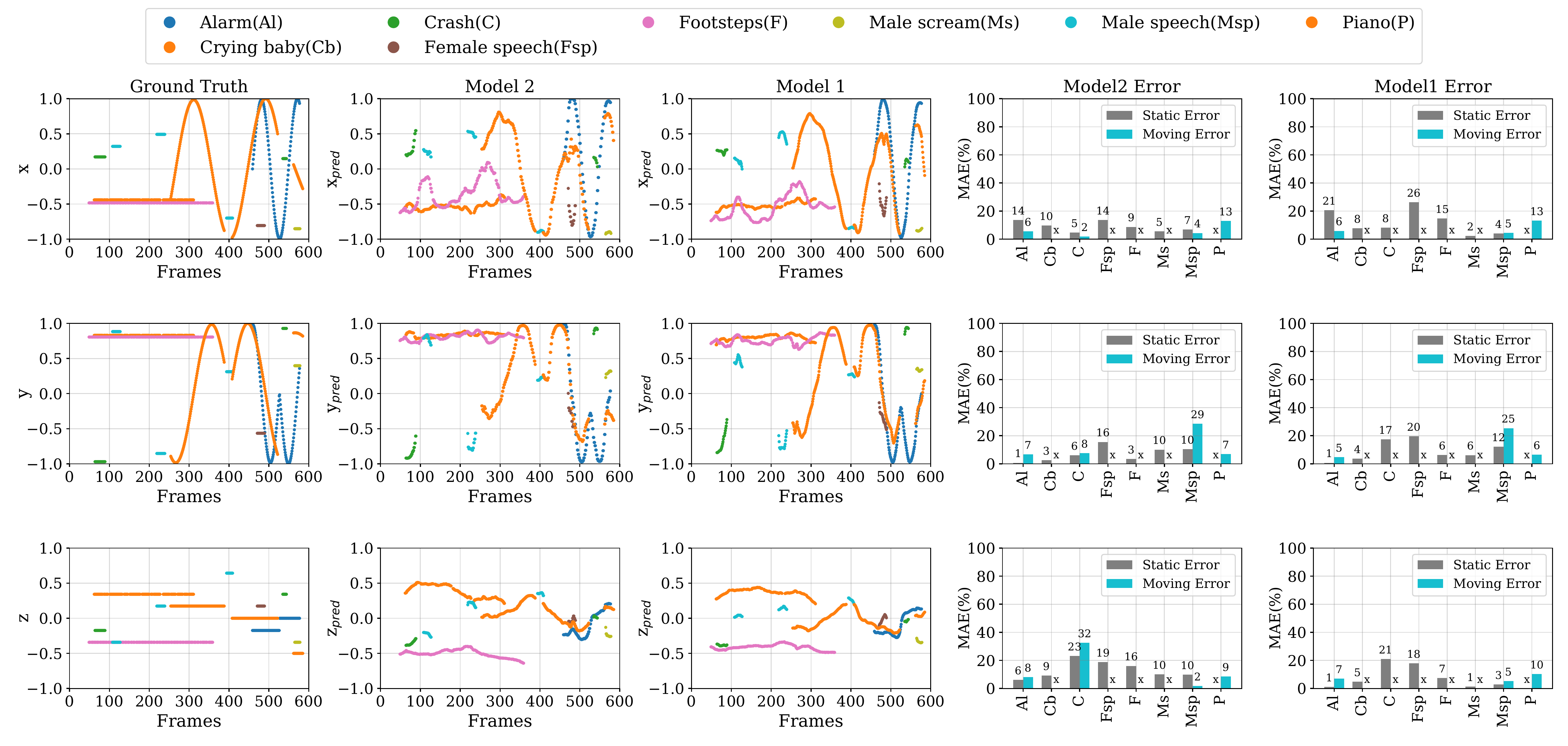}
		\vspace{-0.2cm}
	\caption{Effect of transfer learning: true and predicted trajectories from Model1 and Model2 with classwise MAE.}
	\label{fig:TL_effect}
\end{figure*}

To speed up the training process and reduce the risk of overfitting, we repeated the experiments using the pre-trained CRNN weights from an existing SELD model using SALSA-Lite, where the best model was obtained at the 47th epoch \cite{2022salsa-lite}. The dataset, architecture, and framework for the pre-trained model detailed in \cite{2022salsa-lite} are same as the CRNN body used in this paper. By keeping the CRNN body's weights fixed using the pre-trained SELD model, the overall $P_d$ is increased by 10 \% for both Model1 and Model2 as shown in Table \ref{tab:TL}. From Fig. \ref{fig:TL_effect}, it can be observed that Model1 outperforms Model2 with higher $P_d$ and lower DOA error.\looseness=-1 
\begin{table}
\footnotesize
\begin{tabular}{|l|l|l|l|l|l|l|l|}
\hline
\textbf{SNR} & \textbf{Model} & \textbf{TPs} & \textbf{TPm} & \textbf{FNs} & \textbf{FNm} & \textbf{Pds} & \textbf{Pdm} \\ \hline
\multirow{2}{*}{Clean} & Model1 & 28843
 & 21523  & 13056 & 9892 & 68.8 & 68.5 \\ \cline{2-8} 
 & Model2 &  27637 & 21389 & 14262 & 10026 & 65.9
 & 68 \\ \hline
\multirow{2}{*}{-2dB} & Model1 & 17169 & 13097 & 24730  & 18318 & 40.9 & 41.7 \\ \cline{2-8} 
 & Model2 & 17199  & 12487 & 24700  & 18928 & 41 & 39.7 \\ \hline
\multirow{2}{*}{-5dB} & Model1 & 15812 & 10919 & 26087 & 20496  & 37.7 & 34.7 \\ \cline{2-8} 
 & Model2 & 14136 & 10522 & 27763 & 20893 & 33.7 & 33.4 \\ \hline
\end{tabular}
\caption{Performance of Model1 and Model2 at different SNR (averaged over test data).}
\label{tab:SNRs}
\end{table}
\vspace{-0.3cm}
\subsection{Effect of low SNR levels}
\begin{figure*}
    \centering
	\includegraphics[width=0.8\textwidth]{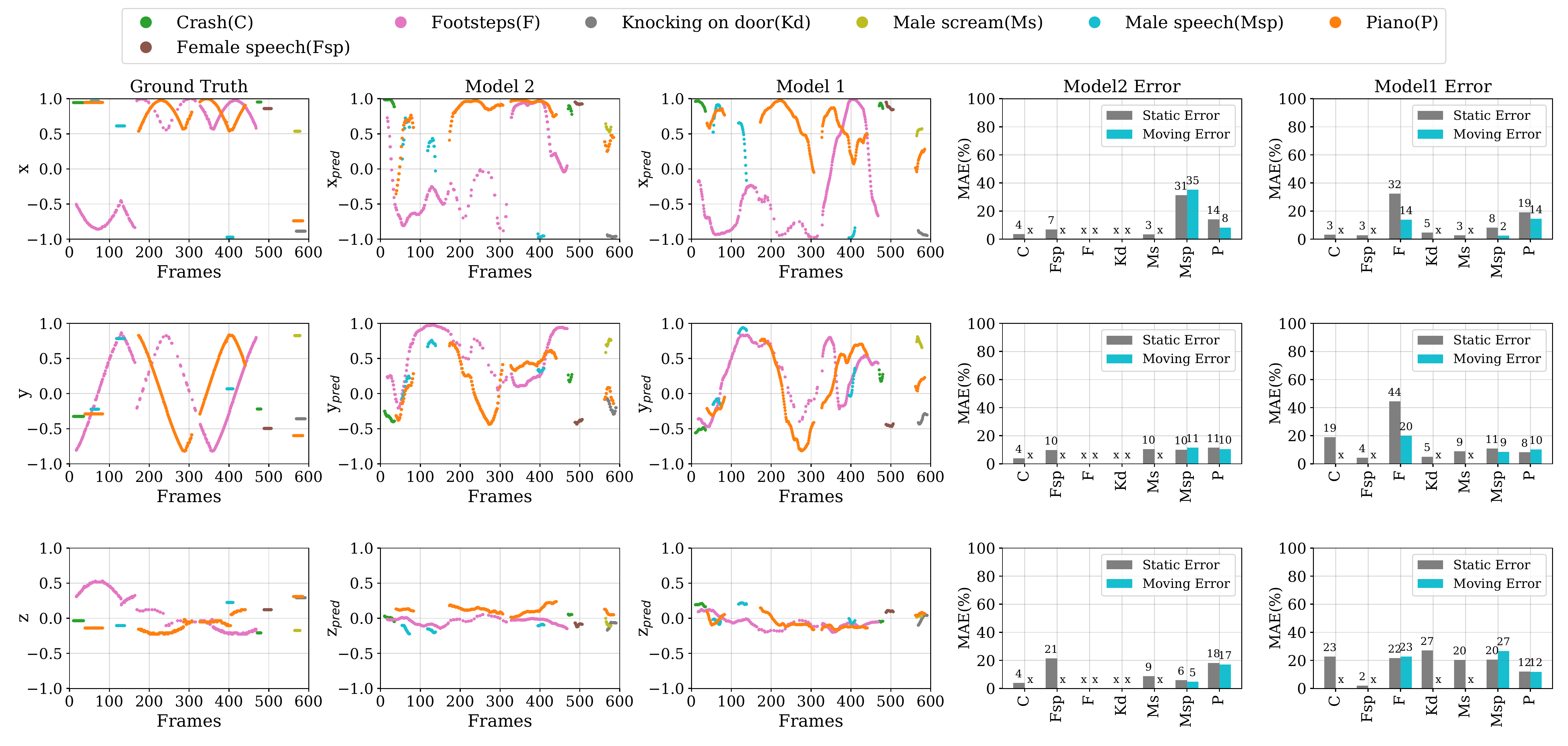}
	\vspace{-0.2cm}
	\caption{Effect of low SNR levels: true and predicted trajectories from Model1 and Model2 with classwise MAE.}
	\label{fig:SNR_effect}
\end{figure*}

Although the dataset consists of unknown interference and noise, to analyze the robustness of both the models, we synthetically added additive white gaussian noise to the recordings. From Table \ref{tab:SNRs}, it is observed that as the SNR level decreases, the $P_d$ of both the models degrades drastically. However, it can also be seen that Model1 outperforms Model2 in noisy scenarios as the estimated derivatives help improve the final DOAs. Fig. \ref{fig:SNR_effect} shows the source trajectories obtained from both Model1 and Model2 for comparison. We observe that the estimates obtained from Model1 are more reliable than those from Model2. 


\begin{table}
\footnotesize
\begin{tabular}{|l|l|l|l|l|l|l|l|}
\hline
\textbf{SNR} & \textbf{Model} & \textbf{TPs} & \textbf{TPm} & \textbf{FNs} & \textbf{FNm} & \textbf{Pds} & \textbf{Pdm} \\ \hline
\multirow{2}{*}{Clean} & Model1 & 33033  & 24687 & 8866 & 6728 & 78.8 & 78.5  \\ \cline{2-8} 
 & Model2 & 33431 & 24531 & 8468 & 6884 & 79.7 & 78 \\ \hline
\multirow{2}{*}{-2dB} & Model1 & 18349 & 12249 & 23550 & 19166 & 43.7 & 39 \\ \cline{2-8} 
 & Model2 & 17777 & 11906 & 24122 & 19509 & 42.4 & 37.8 \\ \hline
\multirow{2}{*}{-5dB} & Model1 & 15365 & 10060 & 26534 & 21355 & 36.7 & 32 \\ \cline{2-8} 
 & Model2 & 15404 & 9816 & 26495 & 21599 & 36.7 & 31.2 \\ \hline
\end{tabular}
\caption{Performance of Model1 and Model2 using the pretrained CRNN SELD model at different SNR (averaged over test data).}
\label{tab:TL}
\vspace{-0.3cm}
\end{table}
\vspace{-0.3cm}
\section{CONCLUSION AND FUTURE WORK}
This work demonstrates that predicting DOA derivatives along with DOAs (Model1) improves the overall localization performance compared to only predicting DOAs (Model2). We show that the proposed Model1 is robust to noise and outperforms Model2 at low SNR levels. 
As the SELD tasks have numerous applications, this analysis shows that estimating DOAs and their derivatives help to improve the source trajectories and overall performance cumulatively. In future, it would be interesting to see the effect of combining higher-order derivatives in the SELD tasks when detection and localization are done simultaneously.            

\small
\bibliographystyle{IEEEbib}
\bibliography{refs}

\end{document}